\documentclass[preprint,12pt]{elsarticle}

 \usepackage{graphics}

\usepackage{amssymb}

\journal{Nuclear Physics A}

\begin{document}

\begin{frontmatter}



\title{On the sensitivity of the energy of vanishing flow towards mass asymmetry of colliding nuclei}
\author{Supriya Goyal}
\author{Rajeev K. Puri\corref{cor1}}
\ead{rkpuri@pu.ac.in}
\cortext[cor1]{Corresponding author}
\address{Department of Physics, Panjab University, Chandigarh, 160 014, India.}

\begin{abstract}
We demonstrate the role of the mass asymmetry in the energy of
vanishing flow by studying asymmetric reactions throughout the
periodic table and over entire colliding geometry. Our results,
which are almost independent of the system size and as well as of
the colliding geometries indicate a sizable effect of the
asymmetry of the reaction on the energy of vanishing flow.
\end{abstract}

\begin{keyword}
heavy-ion collisions \sep multifragmentation \sep quantum
molecular dynamics (QMD) model \sep energy of vanishing flow \sep
mass asymmetric reactions
\end{keyword}

\end{frontmatter}

\section{Introduction}

In the search of nuclear equation of state as well as of nuclear
interactions and forces, collective flow has been found to be of
immense importance \cite{1,2,3,4}. Among collective flow,
transverse in-plane flow enjoys special status. During the last
two decades, much emphasis has been put on the study of collective
flow \cite{1,2,3,4}. Lots of experiments have been performed and
number of theoretical attempts have also been employed to explain
and understand these observations. As reported by \cite{5} for the
first time and later on by many others, collective flow is
negative at low incident energies whereas it is positive at a
reasonable higher incident energies. At a particular incident
energy, however, a transition occurs. This transition energy is
also known as energy of vanishing flow or balance energy. This
energy of vanishing flow (EVF) has been subjected to intensive
theoretical calculations using variety of equations of state as
well as nucleon-nucleon cross-sections \cite{1,2,3,4,5}. This also
includes the mass dependence of EVF which have been reproduced
successfully by various theoretical models \cite{1,3,4}.
\par
Interestingly, most of these studies take symmetric or nearly
symmetric reactions into account. Recently, FOPI group studied the
flow for the asymmetric reaction of $^{40}Ca+^{197}Au$ \cite{6}.
They noted that the flow in the asymmetric collisions is a key
observable for investigating the reaction dynamics. In contrast to
symmetric collisions, where center-of-mass is one of the nucleus,
this quantity is not known {\it a priori} in asymmetric nuclei
experimentally. Later on, FOPI conducted experiment on $^{58}Ni$
and $^{208}Pb$ \cite{6}. In other class of studies, the total mass
of the system was kept fixed as 96 units whereas charge was varied
\cite{ru}. Theoretically, recently Kaur and Kumar \cite{7}
conducted a complete study of the multifragmentation by varying
the asymmetry of the colliding nuclei. Asymmetry parameter $\eta$
is defined as ($A_{T}-A_{p}$/$A_{T}+A_{P}$); where $A_{T}$ and
$A_{P}$ are the masses of the target and projectile, respectively.
All these attempts point towards a need for the systematic study
of the disappearance of flow for asymmetric colliding nuclei.
Further as noted, asymmetry of a reaction plays dramatic role in
heavy-ion collisions \cite{8}. This happens because excitation
energy in symmetric colliding nuclei leads to larger compression
while asymmetric reactions lack the compression since large part
of the excitation energy is in the form of thermal energy.
\par
Note that some isolated studies with asymmetric nuclei are already
done in the literature where the reactions of $^{20}Ne+^{12}C$,
$^{20}Ne+^{27}Al$, $^{20}Ne+^{63}Cu$, $^{58}Ni+^{12}C$,
$^{64}Zn+^{27}Al$, $^{1}H+^{197}Au$, $^{12}C+^{197}Au$,
$^{197}Au+^{12}C$, $^{40}Ar+^{207}Pb$ etc are taken into account
\cite{6,9,10}. Interestingly, none of the studies focus on the
energy of vanishing flow in asymmetric colliding nuclei. To
address this, we here present a systematic study of the EVF as a
function of asymmetry of the colliding nuclei. While total mass of
the reactions remain fixed, the asymmetry is varied by
transferring the neutrons/protons from one nucleus to other. This
study is performed with the quantum molecular dynamics (QMD) model
which is explained briefly in section 2.

\section{The quantum molecular dynamics model}

The quantum molecular dynamics model (QMD) simulates the reaction
on an event-by-event basis \cite{2}. This is based on a molecular
dynamics picture where nucleons interact via two- and three-body
interactions. The explicit two- and three-body interactions
preserve the fluctuations and correlations which are important for
{\it N}-body phenomenon such as multifragmentation \cite{2}.
\par
In the model, the (successfully) initialized nuclei are boosted
towards each other with proper center-of-mass velocity using
relativistic kinematics. Here each nucleon ${\it \alpha}$ is
represented by a Gaussian wave packet with a width of $\sqrt{\it
L}$ centered around the mean position {\it $\vec{r}_{\alpha}$}(t)
and mean momentum {\it $\vec{p}_{\alpha}$}(t) \cite{2}:
\begin{equation}
\phi_{\alpha}(\vec{r},\vec{p},t)=\frac{1}{\left(2\pi
L\right)^{3/4}}e^{\left[-\left\{\vec{r}-\vec{r}_{\alpha}(t)\right\}^2/4L\right]}
e^{\left[i\vec{p}_{\alpha}(t)\cdot\vec{r}/\hbar\right]}.
\end{equation}
The Wigner distribution of a system with ${\it A_{T}+A_{P}}$
nucleons is given by
\begin{equation}
f(\vec{r},\vec{p},t)=\sum_{\alpha =
1}^{A_{T}+A_{P}}\frac{1}{\left(\pi \hbar\right)^{3}}
e^{\left[-\left\{\vec{r}-\vec{r}_{\alpha}(t)\right\}^2/2L\right]}
e^{\left[-\left\{\vec{p}-\vec{p}_{\alpha}(t)\right\}^2
2L/\hbar^{2}\right]^{'}},
\end{equation}
with L = 1.08 $fm^{2}$.
\par The center of each Gaussian (in the
coordinate and momentum space) is chosen by the Monte Carlo
procedure. The momentum of nucleons (in each nucleus) is chosen
between zero and local Fermi momentum
[$=\sqrt{2m_{\alpha}V_{\alpha}(\vec{r})}$; $V_{\alpha}(\vec{r})$
is the potential energy of nucleon $\alpha$]. Naturally, one has
to take care that the nuclei, thus generated, have right binding
energy and proper root mean square radii.
\par
The centroid of each wave packet is propagated using the classical
equations of motion \cite{2}:
\begin{equation}
\frac {d\vec{r}_{\alpha}}{dt} = \frac {dH}{d\vec{p}_{\alpha}},
\end{equation}
\begin{equation}
\frac {d\vec{p}_{\alpha}}{dt} = -\frac {dH}{d\vec{r}_{\alpha}},
\end{equation}
where the Hamiltonian is given by
\begin{equation}
H=\sum_{\alpha} \frac {\vec{p}_{\alpha}^{2}}{2m_{\alpha}} + V
^{tot}.
\end{equation}
 Our total interaction potential $V^{tot}$ reads as \cite{2}
\begin{equation}
V^{tot} = V^{Loc} + V^{Yuk} + V^{Coul} + V^{MDI},
\end{equation}
with
\begin{equation}
V^{Loc} = t_{1}\delta(\vec{r}_{\alpha}-\vec{r}_{\beta})+
t_{2}\delta(\vec{r}_{\alpha}-\vec{r}_{\beta})
\delta(\vec{r}_{\alpha}-\vec{r}_{\gamma}),
\end{equation}
\begin{equation}
V^{Yuk}=t_{3}e^{-|\vec{r}_{\alpha}-\vec{r}_{\beta}|/m}/\left(|\vec{r}_{\alpha}-\vec{r}_{\beta}|/m\right),
\end{equation}
with ${\it m}$ = 1.5 fm and $\it{t_{3}}$ = -6.66 MeV.
\par
The static (local) Skyrme interaction can further be parametrized
as:
\begin{equation}
U^{Loc}=\alpha\left(\frac{\rho}{\rho}_o\right)+
\beta\left(\frac{\rho}{\rho}_o\right)^{\gamma}.
\end{equation}
Here $\alpha, \beta$ and $\gamma$ are the parameters that define
equation of state. The momentum dependent interaction is obtained
by parameterizing the momentum dependence of the real part of the
optical potential. The final form of the potential reads as
\begin{equation}
U^{MDI}\approx t_{4}ln^{2}[t_{5}({\it\vec{p}_{\alpha}}-{\it
\vec{p}_{\beta}})^{2}+1]\delta({\it \vec{r}_{\alpha}}-{\it
\vec{r}_{\beta}}).
\end{equation}
Here ${\it t_{4}}$ = 1.57 MeV and ${\it t_{5}}$ = 5 $\times
10^{-4} MeV^{-2}$. A parameterized form of the local plus momentum
dependent interaction (MDI) potential (at zero temperature) is
given by
\begin{equation}
U=\alpha \left({\frac {\rho}{\rho_{0}}}\right) + \beta
\left({\frac {\rho}{\rho_{0}}}\right)+ \delta
ln^{2}[\epsilon(\rho/\rho_{0})^{2/3}+1]\rho/\rho_{0}.
\end{equation}
The parameters $\alpha$, $\beta$, and $\gamma$ in above Eq. (11)
must be readjusted in the presence of momentum dependent
interactions so as to reproduce the ground state properties of the
nuclear matter. The set of parameters corresponding to different
equations of state can be found in Ref. \cite{2}.

\section{Results and discussion}

For the present study, we simulated various reactions for
1000-5000 events in the incident energy range between 90 and 350
MeV/nucleon. In particular, we simulated the reactions of
$^{17}_{8}O+^{23}_{11}Na$ ($\eta = 0.1$),
$^{14}_{7}N+^{26}_{12}Mg$ ($\eta = 0.3$),
$^{10}_{5}B+^{30}_{14}Si$ ($\eta = 0.5$), and
$^{6}_{3}Li+^{34}_{16}S$ ($\eta = 0.7$) for total mass = 40,
$^{36}_{18}Ar+^{44}_{20}Ca$ ($\eta = 0.1$),
$^{28}_{14}Si+^{52}_{24}Cr$ ($\eta = 0.3$),
$^{20}_{10}Ne+^{60}_{28}Ni$ ($\eta = 0.5$), and
$^{10}_{5}B+^{70}_{32}Ge$ ($\eta = 0.7$) for total mass = 80,
$^{70}_{32}Ge+^{90}_{40}Zr$ ($\eta = 0.1$),
$^{54}_{26}Fe+^{106}_{48}Cd$ ($\eta = 0.3$),
$^{40}_{20}Ca+^{120}_{52}Te$ ($\eta = 0.5$), and
$^{24}_{12}Mg+^{136}_{58}Ce$ ($\eta = 0.7$) for total mass = 160,
and $^{108}_{48}Cd+^{132}_{56}Ba$ ($\eta = 0.1$),
$^{84}_{38}Sr+^{156}_{66}Dy$ ($\eta = 0.3$),
$^{60}_{28}Ni+^{180}_{74}W$ ($\eta = 0.5$), and
$^{36}_{18}Ar+^{204}_{82}Pb$ ($\eta = 0.7$) for total mass = 240.
The present study is for semi-central collisions (i.e. b/b$_{max}$
= 0.25). Note that in some cases, slight variation can be seen for
charges. The charges are chosen in a way so that colliding nuclei
are stable nuclides. A soft equation of state with isotropic
standard energy dependent cugnon cross-section (labeled as
Soft$^{iso}$) and momentum dependent soft equation of state with
standard energy dependent cugnon cross-section (labeled as SMD)
are used in the present reactions.
\par
The energy of vanishing flow (EVF) is calculated using the {\it
directed transverse momentum $<P^{dir}_{x}>$}, which is defined
as:
\begin{equation}
\langle P_{x}^{dir}\rangle=\frac{1}{A}\sum_i {\rm
sign}\{Y(i)\}~{\bf{p}}_{x}(i),
\end{equation}
where $Y(i)$ and ${\bf{p}}_{x}(i)$ are the rapidity distribution
and transverse momentum of $i^{th}$ particle, respectively.
\par
In Fig. 1, we display the time evolution of the directed
transverse flow $<P^{dir}_{x}>$, for various asymmetric reactions
with system mass (sysmass) = 80 and 240 units. From the figure, it
is clear that transverse flow saturates quite early during the
interaction. A negative value during the initial stages signifies
the dominance of the attractive interactions. One also sees
well-known trends of transverse collective flow with increase in
the incident energy. A clear transition from the negative flow to
positive flow is also visible. From the figure, one can see that
nearly symmetric reactions ($\eta$ = 0.1 and 0.3) respond strongly
to the change with incident energy compared to highly asymmetric
reactions. One notices that for $\eta$=0.1 in the system mass 240,
the change in the flow between 40 and 400 MeV/nucleon is 50 MeV,
whereas for the same energy range, the flow varies between -2 MeV
and +18 MeV (net 20 MeV) for $\eta$=0.7. The cause behind is that
with the increase in the asymmetry of a reaction lesser binary
collisions take place resulting in lesser density and therefore,
less response occurs for variation in the collective flow. On the
other hand, for nearly symmetric reactions, the binary collisions
increase linearly with incident energy, therefore, huge difference
can be seen.
\par
It would be interesting to see how mass dependence character of
the EVF behaves at a fixed asymmetry. In Fig. 2, we display the
EVF verses combined system mass by keeping the asymmetry fixed. We
notice that in all the cases, a perfect power law dependence (with
power factor close to 1/3) can be seen for all asymmetries right
from 0.1 to 0.7. All points lie on the line indicating that
asymmetry plays a major role in the mass dependence. For a given
mass e.g. system mass = 80 units, we see a variation of 50 MeV in
EVF for $\eta$ varying between 0.1 and 0.7. This also explains why
in the earlier mass dependence studies \cite{1}, though average
behavior was a power law, individual EVFs were quite far from the
average law \cite{1,3}. There this happened because no control was
made for the asymmetry of a colliding pair. Had all reactions been
analyzed on a fixed asymmetry, one could have obtain perfect match
with average power law. Further, a careful comparison of Soft and
SMD reveals that the difference between EVF with both methods
narrows down for light nuclei with larger asymmetry. In another
calculations at b/b$_{max}$ = 0.5 (not shown here), one noticed
that MDI gives smaller EVF for lighter nuclei whereas EVF with MDI
is enhanced for heavier nuclei. The difference is clearly visible
for larger asymmetries. The cause of this behavior lies in the
fact that MDI are more attractive compared to soft static equation
of state at low densities whereas they turn more repulsive at
higher incident energies. Based on this information, it is clear
that for lighter nuclei at large asymmetry (and therefore, low
densities) MDI suppresses the EVF. These observations are more
visible at large asymmetries.
\par
In Fig. 3, we display EVF as a function of $\eta$ for a fixed mass
equal to 40, 80, 160, and 240 units. The results using the
Soft$^{iso}$ (top panel) and SMD (bottom panel) EoS are displayed
for clarity. In agreement with all previous studies, EVF decreases
with increase in the mass of the system with both Soft$^{iso}$ and
SMD EoS. This decrease has been attributed to the increasing role
of Coulomb forces in heavier colliding nuclei. As discussed in
earlier figures, a sizable influence can be seen towards EVF with
variation of the asymmetry of a reaction. For lighter masses, the
effect of the variation of $\eta$ can result about 40 MeV change
in the EVF. As noted in absolute terms, lighter nuclei are more
affected compared to heavier ones. Overall, one sees that the
effect of the asymmetry of a reaction is not at all negligible. It
can have sizable effect which goes as power law with power factor
close to 1/3.
\par
In Fig. 4, we display the percentage difference $\Delta$EVF(\%)
defined as $\Delta$EVF(\%) =
(($EVF^{\eta\neq0}$-$EVF^{\eta=0}$)/$EVF^{\eta=0}$)$\times$100.
Very interestingly, we see that the effect of the asymmetry
variation is almost uniform throughout the periodic table. The
mean variation is 0.7\%, 4\%, 13\%, and 39\% with Soft$^{iso}$ EoS
and 0.3\%, 2\%, 6\%, and 20\% with MDI for $\eta$ = 0.1, 0.3, 0.5,
and 0.7, respectively. In other words, asymmetry of a reaction can
play significant role in EVF and the deviation from the mean line
can be eliminated if proper care is taken for the asymmetry of
colliding nuclei.
\par
In Fig. 5, we display the EVF as a function of the impact
parameter for different asymmetries. A well known trend i.e.
increase in the EVF with impact parameter can be seen. Further, as
demonstrated by many authors, the impact parameter variation is
less effected in the presence of MDI. The striking result is that
the effect of mass asymmetry variation is almost independent of
the impact parameter.

\section{Summary}

Using the quantum molecular dynamics model, we presented a
detailed study of the balance energy with reference to mass
asymmetry. Almost independent of the system mass as well as impact
parameter, a uniform effect of the mass asymmetry can be seen at
the energy of vanishing flow. We find that for large asymmetries
($\eta$ = 0.7), the effect of asymmetry can be 15\% with MDI and
in the absence it can be 40\%. This also explains the deviation in
the individual EVF from the mean values as reported earlier.

\section{Acknowledgement}

The work is supported by a grant from the Department of Science
and Technology (DST), Govt. of India via grant no.
SR/S2/HEP-28/2006.
\bibliographystyle{elsarticle-num}

\newpage
{\Large \bf Figure Captions}\\

{\bf FIG. 1.} (Color Online) The time evolution of the directed
transverse flow $<P^{dir}_{x}>$ as a function of the reaction time
using different mass asymmetries. The left panel is for the
reaction leading to sysmass = 80 units whereas right panel
indicates sysmass = 240 units. The results for different mass
asymmetries $\eta$ = 0.1, 0.3, 0.5, and 0.7 are represented,
respectively, by the solid, dashed, dotted, and dashed-dotted
lines.\\

{\bf FIG. 2.} (Color Online) The EVF as a function of total mass
(sysmass) of reacting partners. The upper panel is for
Soft$^{iso}$ EoS whereas lower panel is for SMD. The results for
different asymmetries $\eta$ = 0.1, 0.3, 0.5, and 0.7 are
represented, respectively, by the solid squares, circles,
triangles and inverted triangles. Lines are power law fit
$\propto$ $A^\tau$.\\

{\bf FIG. 3.} (Color Online) The EVF as a function of asymmetry
parameter $\eta$. The upper panel is for Soft$^{iso}$ EoS whereas
lower panel is for SMD. The results for the systems having total
mass A of 40, 80, 160, and 240 are represented, respectively, by
the open squares, circles, triangles and inverted triangles for
the case of SMD whereas shaded area in the case of Soft$^{iso}$
EoS represents the effect of charge variation leading to fixed
value of charge and mass of the reaction. Lines are to guide the
eye.\\

{\bf FIG. 4.} (Color Online) The percentage difference
$\Delta$EVF(\%) as a function of sysmass of reacting partners. The
upper panel is for Soft$^{iso}$ EoS whereas lower panel is for
SMD. The results of the percentage difference for different
asymmetries $\eta$ = 0.1, 0.3, 0.5, and 0.7 are represented,
respectively, by the solid squares, circles, triangles and
inverted triangles. Horizontal lines represent the mean value of
$\Delta$EVF(\%) for each $\eta$.\\

{\bf FIG. 5.} (Color Online) The EVF as a function of reduced
impact parameter for the sysmass 240. The upper panel is for
Soft$^{iso}$ EoS whereas lower panel is for SMD. The results of
the percentage difference for different asymmetries $\eta$ = 0.1,
0.3, 0.5, and 0.7 are represented, respectively, by the solid
squares, circles, triangles and inverted triangles. Lines are to
guide the eye.\\
\begin{figure}[!tb]
\centering \vskip -1.8cm \setlength{\abovecaptionskip}{-10cm}
\setlength{\belowcaptionskip}{0.5cm}
\includegraphics[width=13.8cm]{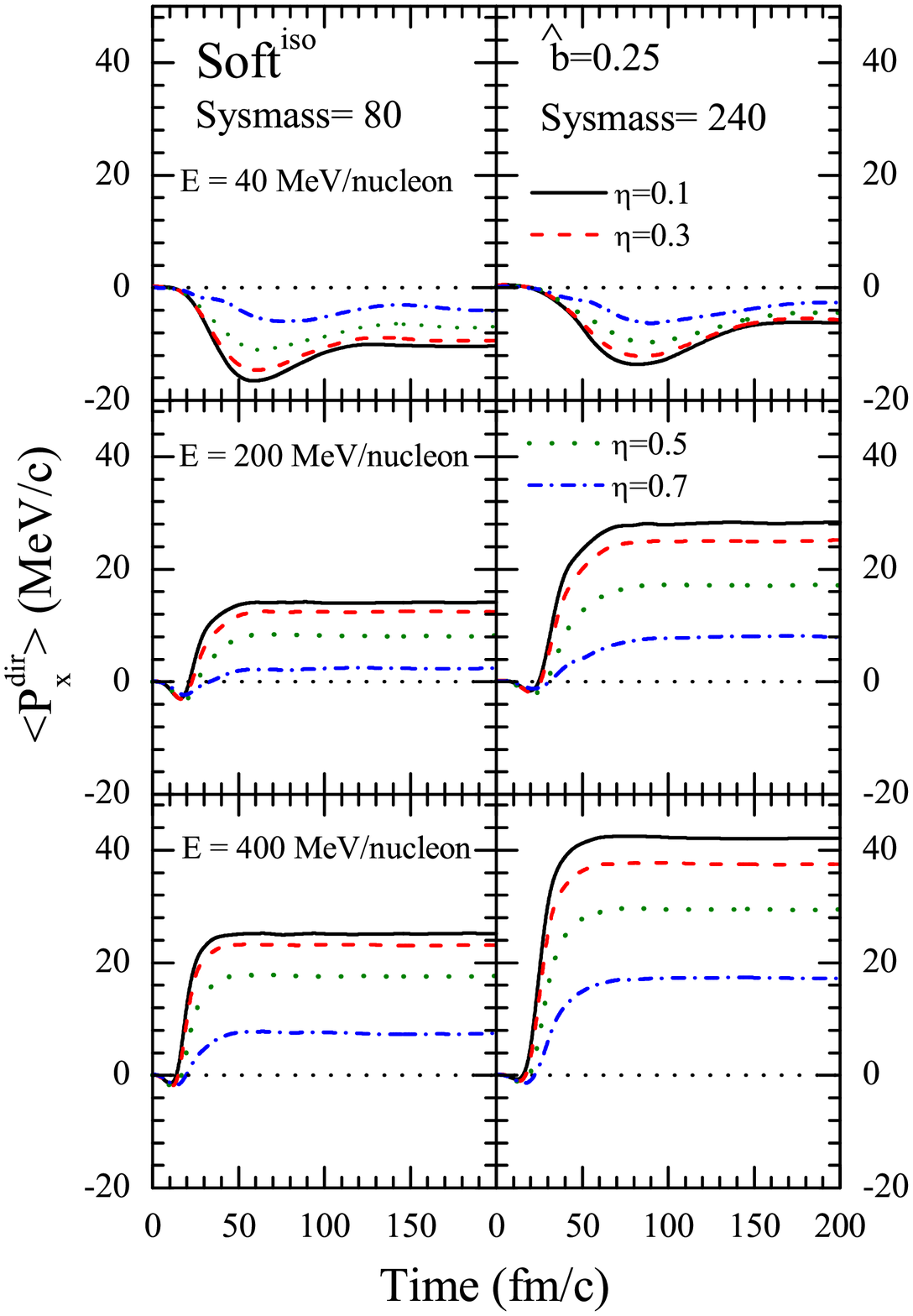}
\end{figure}
\begin{figure}[!tb]
\centering \vskip -1.8cm \setlength{\abovecaptionskip}{-10cm}
\setlength{\belowcaptionskip}{0.5cm}
\includegraphics[width=13.8cm]{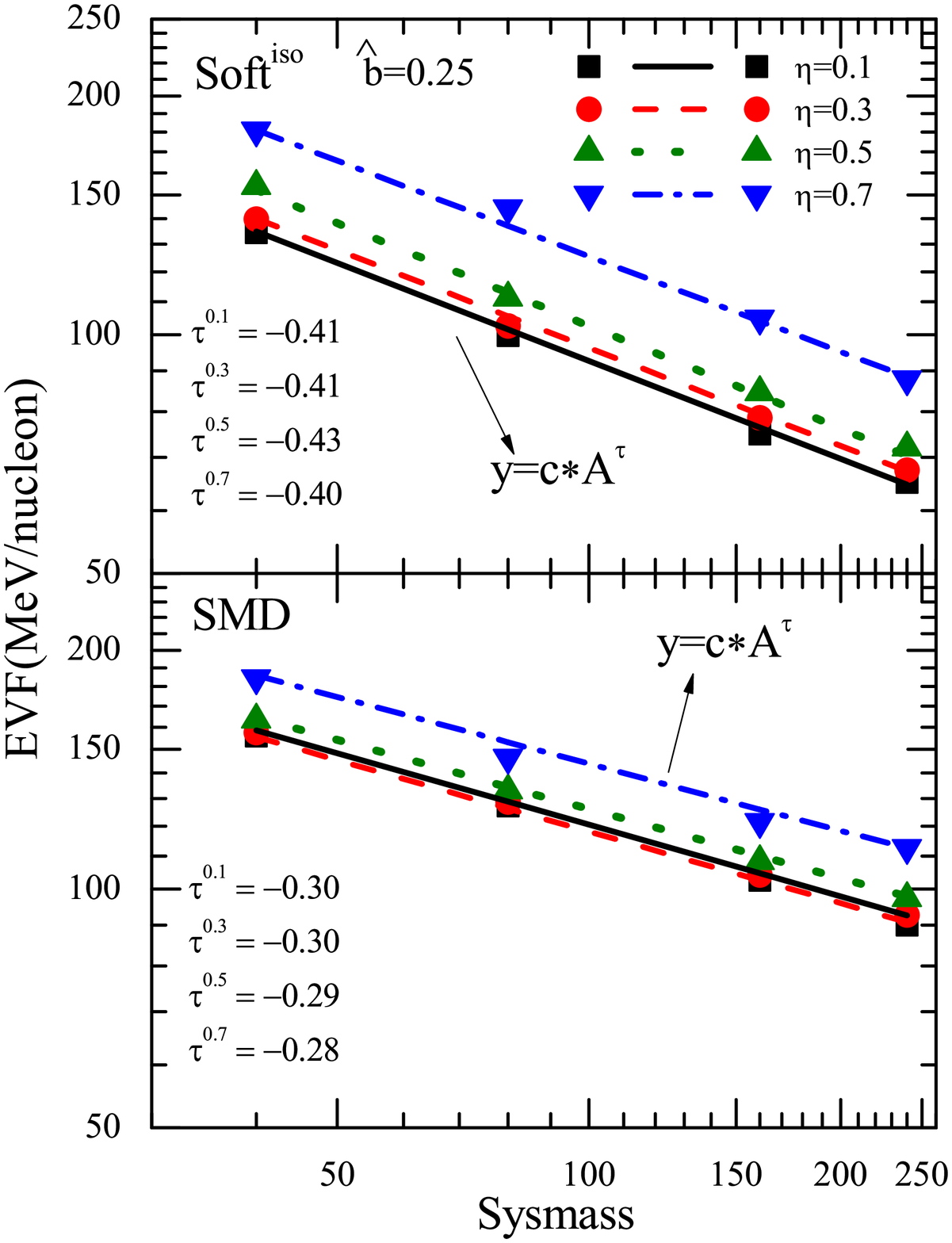}
\end{figure}
\begin{figure}[!tb]
\centering \vskip -1.8cm \setlength{\abovecaptionskip}{-10cm}
\setlength{\belowcaptionskip}{0.5cm}
\includegraphics[width=13.8cm]{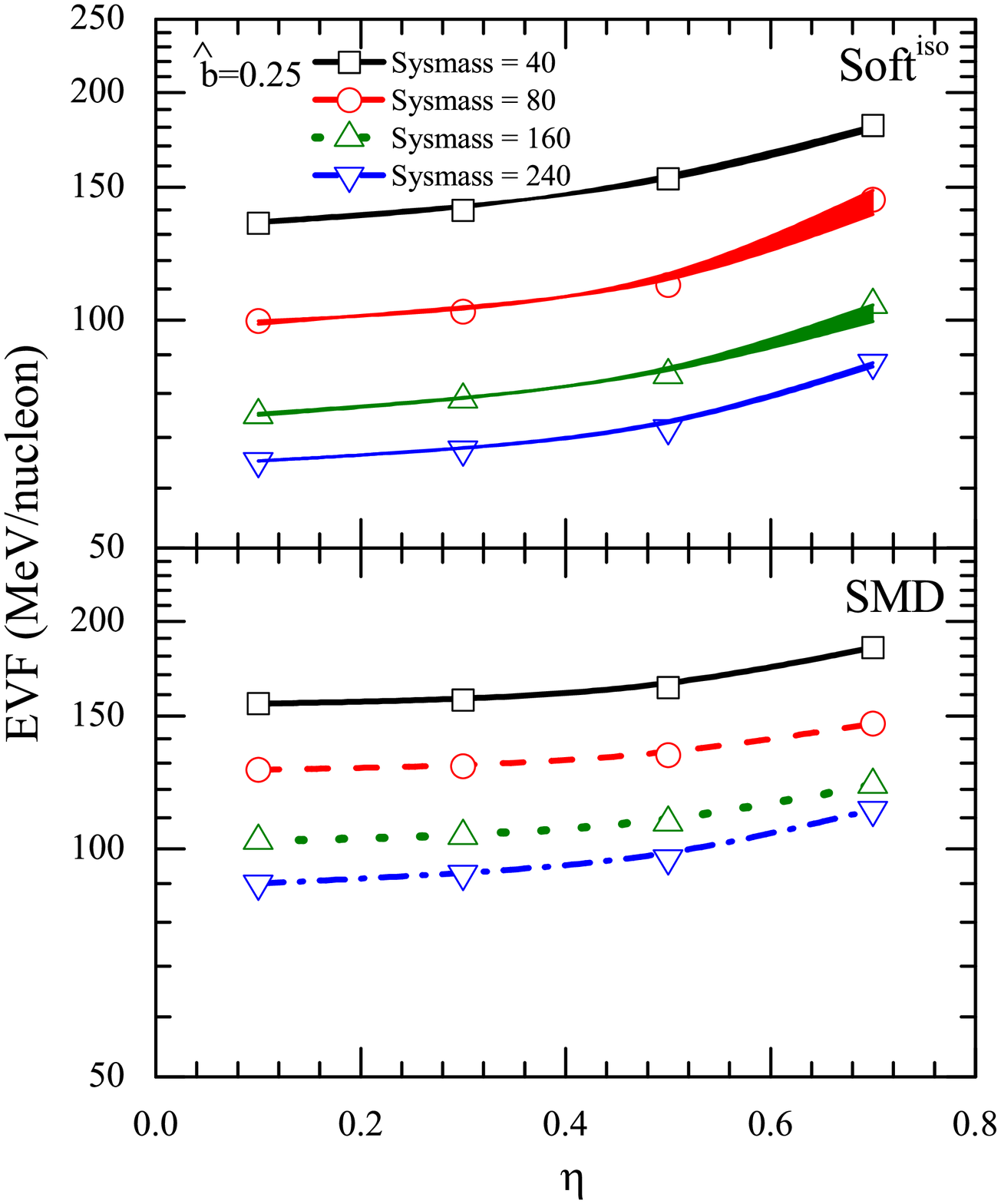}
\end{figure}
\begin{figure}[!tb]
\centering \vskip -1.8cm \setlength{\abovecaptionskip}{-10cm}
\setlength{\belowcaptionskip}{0.5cm}
\includegraphics[width=13.8cm]{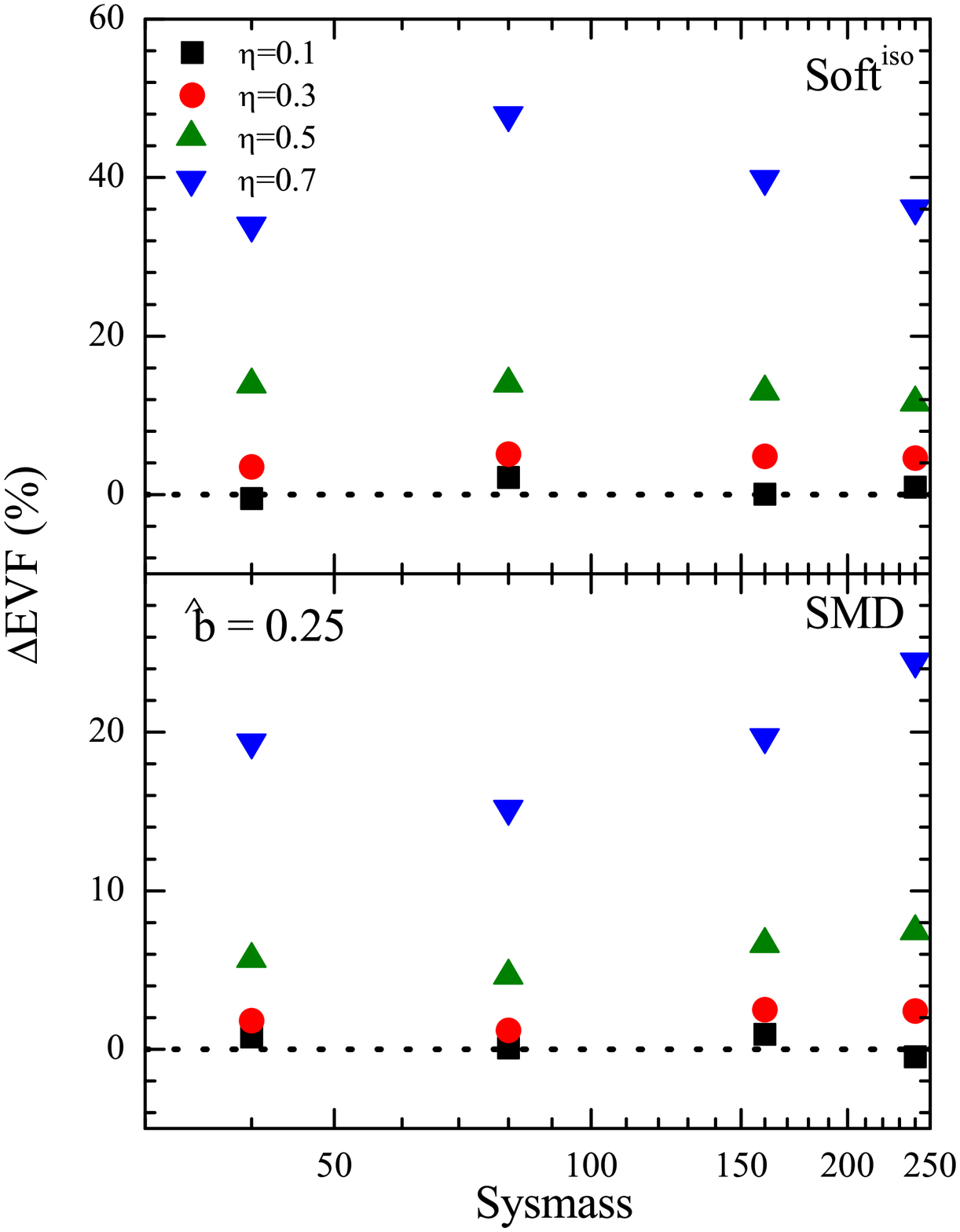}
\end{figure}
\begin{figure}[!tb]
\centering \vskip -1.8cm \setlength{\abovecaptionskip}{-10cm}
\setlength{\belowcaptionskip}{0.5cm}
\includegraphics[width=13.8cm]{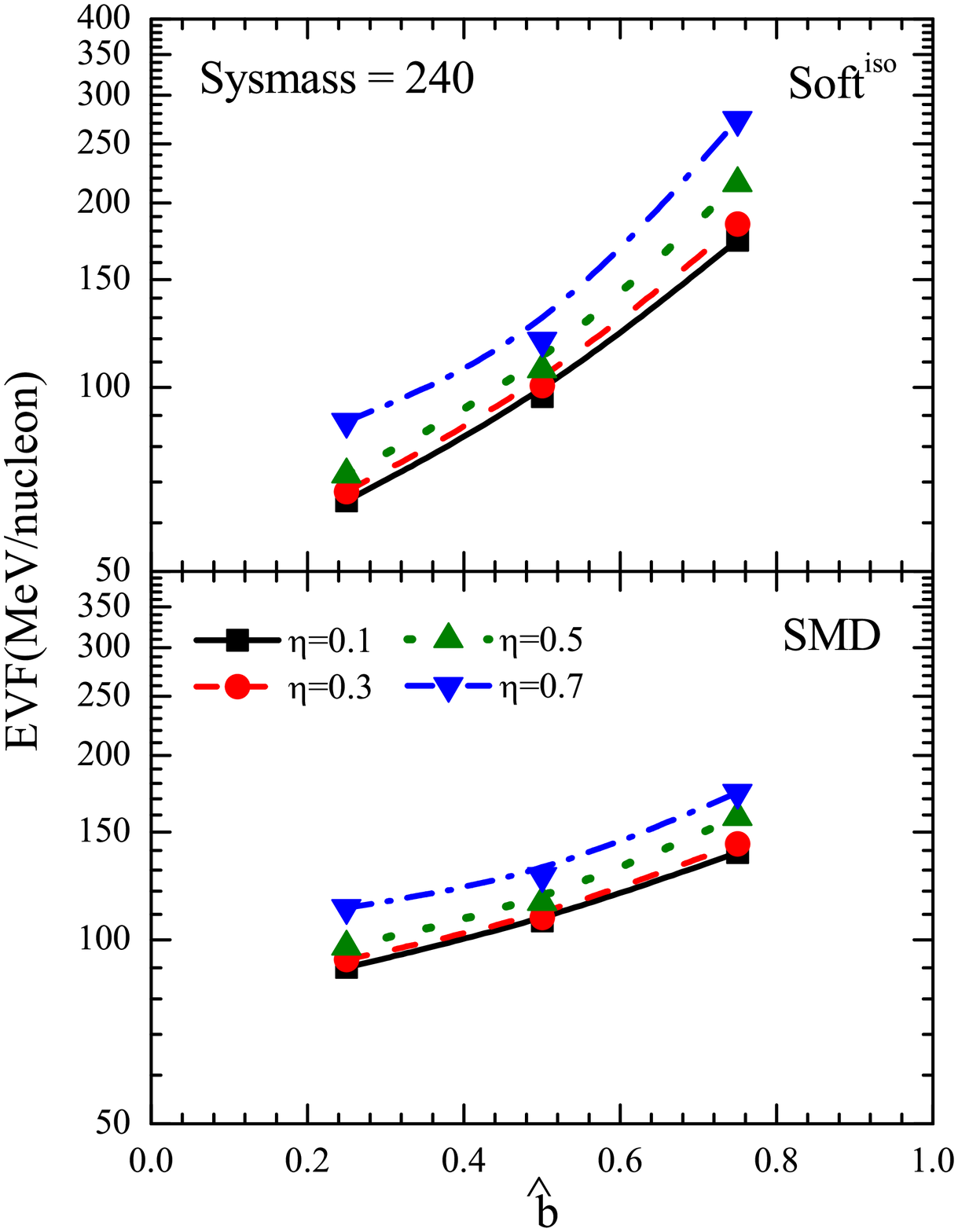}
\end{figure}
\end{document}